\documentclass{elsart5}

 \usepackage{graphicx}

\usepackage{amssymb}
\def\pr{PrOs$_4$Sb$_{12}$}

\begin{document}

\begin{frontmatter}

\title{Superconducting gap nodes in PrOs$_4$Sb$_{12}$}

\author[aff1]{S. H. Curnoe}
\ead{curnoe@physics.mun.ca}
\author[aff1]{T. R. Abu Alrub}
\author[aff2]{I. A. Sergienko}
\author[aff3]{I. Vekhter}
\address[aff1]{Department of Physics \& Physical Oceanography,
Memorial University of Newfoundland, St. John's, Canada}
\address[aff2]{Materials Science and Technology Division, Oak Ridge
National Laboratory, Oak Ridge, TN 37831, USA}
\address[aff3]{Department of Physics and Astronomy, Louisiana State University,
Baton Rouge, LA 70803, USA}
\received{12 June 2005}
\revised{13 June 2005}
\accepted{14 June 2005}


\begin{abstract}
We examine the superconducting gap nodes in crystals
with tetrahedral (T$_h$)
symmetry.
The $(0,0,1)$  phase of the three dimensional order parameter
in the triplet channel has nodes in the $[001]$ directions.
Following a second order phase transition to the state
$(0,i|\eta_2|,|\eta_1|)$, each node
lifts away from the Fermi surface and splits into two deep dips.
We discuss this scenario in the context of 
multiple superconducting phases in \pr.

\end{abstract}


\end{frontmatter}

In the years since the discovery of multiple superconducting phases in \pr\
experimental results concerning the 
symmetry of the superconducting  states have proliferated,
but so far no theoretical description is widely accepted.
Our starting point is a strict analysis of
symmetry and symmetry-breaking described by 
Landau theory \cite{ueda,us}.
This approach makes definite statements about the 
symmetry and gap nodes of possible superconducting states.
In this proceedings, we extend this analysis to 
discuss the possible existence of deep dips in the gap function, and 
how they may be observed experimentally.

\pr\ has two superconducting phases, as observed in thermal conductivity \cite{izawa}
and several other measurements
\cite{aoki,chia,grube}, the `A-phase', which borders the normal phase
on the H-T phase diagram
and the `B-phase' which appears just below it.
There are more experimental observations of the B-phase than the A-phase,
in part because it occupies a much greater area of the phase diagram.
Various measurements have 
determined that the B-phase has triplet pairing, breaks time-reversal symmetry
\cite{aoki}
and has nodes or near-nodes in the $[001]$ directions of the gap function
\cite{izawa,chia}.
Some experiments have interpreted the double transition in terms of
multi-band superconductivity \cite{seyfarth}, such that the A-phase and
B-phase possess the same symmetry.
A third superconducting phase has been detected deep in the superconducting region
of the H-T phase diagram,  but its symmetry
properties are presently unknown \cite{cic2005}.

As the starting point of this discussion, we select from Table I of Ref.
\cite{us} the state which best matches the description of the B-phase.
This table is a list of all possible superconducting states for
crystals with tetrahedral symmetry with their corresponding gap nodes.
Broken time-reversal symmetry is indicated by the absence of the element
$K$ in the symmetry groups.
We also take into consideration the second order phase transition sequences,
shown in Figs.\ 1 and 2 of Ref. \cite{us}; the B-phase should be two 
steps away from the normal phase.
The best choice for the B-phase is given by a three
component order parameter, which transforms according to the
3D representation  of the tetrahedral point group, with components
$(0, i|\eta_2|, |\eta_1|)$, in the triplet channel.
This phase breaks time reversal symmetry, and it is uniquely accessible 
both directly from the normal phase and from the phase $(0,0,1)$ by a 
second order phase transition.
However, strictly speaking, this phase is nodeless.
The A-phase is identified with the state $(0,0,1)$. 
The symmetry groups associated with $(0,0,1)$ and $(0, i|\eta_2|, |\eta_1|)$
are $D_2(C_2) \times K$ and $D_2(E)$ respectively \cite{us}.
Each of these includes non-trivial combinations of the $D_2$ point group
operations, gauge elements and/or
time reversal, and a group-subgroup relation exists between them.

Although a lowering of symmetry through the A-B transition is expected,
the four-fold to two-fold symmetry lowering observed
in a thermal conductivity experiment \cite{izawa}
is not described by this scenario.  
The probable existence of domains may account for the four-fold
observation, especially since in the tetrahedral point group there can be no
four-fold symmetry, 
but then it is impossible to account for the two-fold
observation.  Clearly, confirmation of these results would be very useful.

The basis functions for the 3D representation in the triplet channel are
$$
\{\mathbf{d}_1(\mathbf{k}),\mathbf{d}_2(\mathbf{k}),\mathbf{d}_3(\mathbf{k})\}=
$$
$$
\hspace{.2in}
\{a\hat{\mathbf{y}}k_z+b\hat{\mathbf{z}}k_y, a\hat{\mathbf{z}}k_x+b\hat{\mathbf{x}}k_z,a\hat{\mathbf{x}}k_y+b\hat{\mathbf{y}}k_x
\},
$$ where
$a$ and $b$ are real numbers.  Under octahedral symmetry,
$|a|=|b|$.  Since the Fermi surface is approximately octahedral \cite{suga2002},
we will assume $|a|\approx |b|$.
The gap functions are
$$
\Delta_{\pm}(\mathbf k) = [|\mathbf{d}(\mathbf{k})|^2\pm|\mathbf{d}(\mathbf{k})\times
\mathbf{d}^{*}(\mathbf{k})|]^{1/2},
$$
where 
$$
\mathbf{d}(\mathbf{k}) = \sum_i\eta_i \mathbf{d}_i(\mathbf{k})
$$
and $\eta_i$ are the components of the order parameter.
There are 
two different gap functions when $\mathbf{d}^{*}(\mathbf{k}) \neq
\mathbf{d}(\mathbf{k})$, {\em i.e.} when time reversal symmetry is broken.
In that case, usually only the `$-$' sign yields a gap function with nodes.

For the A-phase $(0,0,1)$, we have
$\mathbf{d}(\mathbf{k}) = a\hat{\mathbf{x}}k_y+b\hat{\mathbf{y}}k_x$
with gap 
$[a^2 k_y^2 + b^2 k_x^2]^{1/2}$, which
has two point nodes in the $[001]$ directions, as shown in Fig.\ 1a).  
These nodes are a strict consequence of symmetry and are not an
artifact of the choice of basis functions.
The B-phase $(0,i|\eta_2|,|\eta_1|)$ 
emerges as a result of a second order phase transition, 
so $|\eta_2|$ is small close to the transition line.  
The gap function of the B-phase is
$$
\Delta(\mathbf{k}) = [ 
(|\eta_1|^2 b^2 + |\eta_2|^2 a^2)k_x^2 
+ |\eta_1|^2 a^2 k_y^2
+|\eta_2|^2 b^2 k_z^2  
$$
$$
-2 |\eta_1||\eta_2||k_x|\sqrt{a^2b^2k_x^2+a^4k_y^2+b^4k_z^2}
]^{1/2}.
$$
For small 
$\eta_2$, the nodes of the A-phase lift away from the Fermi surface,
as expected, and split into two, as shown in Fig.\ 1b).  
Larger values of $\eta_2$ cause the
dips to move around with respect to the Fermi surface, and the
dips become less pronounced in general, as shown in Fig.\ 1c).

\begin{figure}[h]
\includegraphics[scale =1]{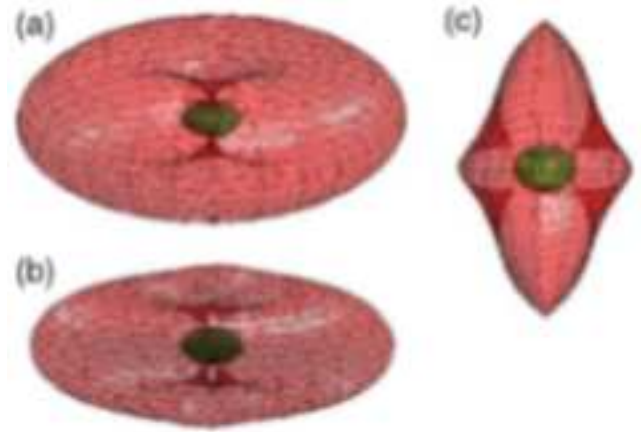}
\caption{Gap functions drawn over a spherical Fermi surface for
$a=1.2$ and $b=1$. a) 
The A-phase, $\eta_1=$1.1, $\eta_2 = 0$ with nodes in the $[0,0,1]$ direction.
b)  The B-phase, $\eta_1=1$, $\eta_2= 0.2$.  c) The B-phase, $\eta_1=1$, $\eta_2 = 0.85$.
}
  	\label{fig-1}
  \end{figure}

If the transition between the A-phase and the B-phase is
second order then the gap minimum in the B-phase will be small
in the vicinity of the phase boundary.
Although in general, finite gaps will destroy
power-law temperature dependencies in specific heat,
{\em etc.}, 
at temperatures large compared to the gap minimum power laws may still be observe.
\cite{huss}.   
As the temperature is
lowered,
two effects are at work, first, measurements become more
sensitive to the finite energy gaps because of the reduced temperature
and, second, the B-phase may evolve
from a state with deep dips in the energy gap 
function and a small gap minimum toward a state in which the dips are
less pronounced. 

Experimentally, the situation is far from clear.  Power
law temperature dependencies are observed in the specific heat
\cite{exper},
thermal conductivity \cite{izawa} and 
penetration depth \cite{chia}. 
On the other hand, nodelessness has been interpreted from
nuclear quadrupole resonance \cite{kotegawa}, $\mu$SR \cite{mclaugh}
and tunneling
spectroscopy \cite{suder}.  Various temperature ranges were studied
in each case.  Each of these methods represents a different approach 
to the detection of quasiparticles in gap nodes.  The
cross-over between universal and non-universal scaling 
described in this proceedings would likely be observed differently
in each experimental arrangement.
To test the scenario put forward in this proceedings,
evidence of a cross-over between
universal and non-universal scaling behaviour should be 
sought within a single set-up.
Of course,
the ideal measurement would be a direct, directional-dependent
detection of nodes or near nodes, especially of the splitting of
nodes through the phase transition.

\vspace{.2in}
This work was supported by NSERC of Canada and by the Board of Regents of 
Louisiana.


\end{document}